\documentclass{iaus}
\usepackage{graphics,graphicx}

   \checkfont{eurm10}
   \iffontfound
     \IfFileExists{upmath.sty}
       {\typeout{^^JFound AMS Euler Roman fonts on the system,
                    using the 'upmath' package.^^J}%
        \usepackage{upmath}}
       {\typeout{^^JFound AMS Euler Roman fonts on the system, but you
                    dont seem to have the}%
        \typeout{'upmath' package installed. iaus.cls can take advantage
                  of these fonts,^^Jif you use 'upmath' package.^^J}%
       }
   \else
   \fi

   \checkfont{msam10}
   \iffontfound
     \IfFileExists{amssymb.sty}
       {\typeout{^^JFound AMS Symbol fonts on the system, using the
                 'amssymb' package.^^J}%
        \usepackage{amssymb}%

       }{}
   \fi

   \IfFileExists{amsbsy.sty}
     {\typeout{^^JFound the 'amsbsy' package on the system, using it.^^J}%
      \usepackage{amsbsy}}
     {}

\newsavebox{\astrutbox}
\sbox{\astrutbox}{\rule[-5pt]{0pt}{20pt}}

\title[The Interplay among Black Holes, Stars and ISM in Galactic
        Nuclei]{Modelling the Pan-Spectral Energy Distributions of Starburst 
\& Active Galaxies}

\author[M.A.~Dopita, {\it et al.\/}]%
{M.A.~Dopita$^1$, J.~Fischera$^1$, B.~Groves$^1$, R.S.~Sutherland$^1$,\break
L.J.~Kewley$^2$, R.~Tuffs$^3$, C.~Popescu$^3$, \and C.~Leitherer$^4$}

\affiliation{$^1$Research School of Astronomy \& Astrophysics, 
Australia National University, Canberra, Australia. email: 
Michael.Dopita@anu.edu.au\\[\affilskip]
$^2$Harvard-Smithsonian Center for Astrophysics, Cambridge MA, 
USA\\[\affilskip]
$^3$MPI f\"ur Kernphysik, Heidelberg, Germany\\[\affilskip]
$^4$Space Telescope Science Institure, Baltimore MD, USA}
\pubyear{2004}
\volume{222}
\pagerange{1--8}
\date{?? and in revised form ??}
\jname{The Interplay among Black Holes, Stars and ISM \\in Galactic Nuclei}
\editors{Th. Storchi Bergmann, L.C. Ho \& H.R. Schmitt, eds.}
\begin{document}

\maketitle

\begin{abstract}
We present results of a self-consistent model of the spectral energy
distribution (SED)
of starburst galaxies. Two parameters control the IR SED, the mean
pressure in the ISM and
the destruction timescale of molecular clouds. Adding a simplified
AGN spectrum provides
mixing lines on IRAS color : color diagrams. This reproduces the observed
colors of both AGNs and starbursts.
\end{abstract}

\firstsection 
\section{SED Modelling}
Our model, which will be described in detail elsewhere (\cite{Dop04}),
combines the
output of the stellar spectral synthesis code STARBURST 99 as input
to the nebular modelling
code MAPPINGS IIIq, including a 1-D dynamical evolution model of HII regions
around massive clusters of young stars. This model includes all
relevant dust and gas
physics to provide the nebular line, continuum and dust re-emission spectrum.
This allows us produce purely theoretical synthetic spectral energy
distributions (SEDs) in the $0.09-1000\mu$m wavelength range.

We have applied this code to make models of solar metallicity
starbursts lasting some $10^8$~years. We find that
the expansion of the H{\sc ii} region ``stalls" when the internal pressure
matches the pressure of the
ambient ISM, c.f. \cite{Oey98}. Dust in high pressure environments is closer
to the exciting stars, and hotter, so that the far-IR SED peaks at
shorter wavelengths. In addition,
the covering factor of the molecular clouds with respect to the
central stars largely determines
the strength of the PAH emission features in the $3-30\mu$m band,
since the PAHs are destroyed
in the diffuse and ionized phases. Molecular clouds around HII regions are
destroyed and dispersed over time. As a result, the starburst SED shape is
largely controlled by this molecular cloud
dissipation timescale which determines the visibility of the central cluster
at UV wavelengths and the strength of the PAH emission bands in the IR.
By contrast, the ambient ISM pressure determines the SED mainly by controlling
the dust temperature, which affects the peak of the FIR dust re-emission
feature.

Our models use a solar abundance set (as modified by the latest abundance
determinations of \cite{Allende02}, and references therein)
depletion factors given by
\cite{Dopita00}. To reproduce the PAH emission features, we require that
$\sim70$\% of the C is locked up in PAHs. As a consequence, when PAHs
are destroyed by photodisscociation,
the 2200\AA\ absorption feature becomes very weak, and the resulting
theoretical attenuation curve
looks almost identical to the \cite{Calzetti01} curve determined
empirically for starbursts.

Here, we have also added an AGN contribution, assumed to be Synchrotron
self-Compton emission from a trapped jet,
becoming Synchrotron self-absorbed in the $10-100\mu$m wavelength
band. For a ``pure" AGN, this model
reproduces the \cite{Dopita98} mean Seyfert I colors, derived from
the \cite{Rush93} IRAS dataset.
Figure 1 shows IRAS color : color diagrams for various AGN
contributions, expressed as a percentage of the
bolometric flux of the starburst.

We can conclude that our models reproduce the range of FIR colors seen in
starburst galaxies, and that objects classified as starbursts may in fact
contain up to 10\%  contribution from an AGN.

\begin{figure}
\includegraphics[width=130mm]{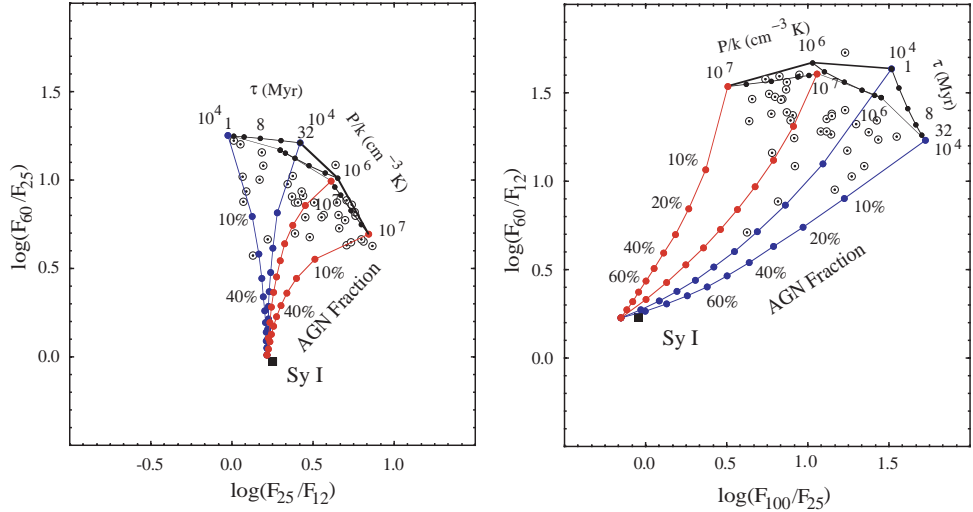}
\caption{IRAS color : color diagrams for the \cite{Rush93}
starbursts compared with our models.
We also show mixing lines with our AGN model. The observed starbursts
may contain up
to 10\% of an AGN contribution.}
\end{figure}

\begin{acknowledgements}
Dopita acknowledges support through a Federation Fellowship.
Dopita, Sutherland \& Fishera acknowlege ARC
Discovery project DP0208445. Kewley is supported by a
Harvard-Smithsonian CfA Fellowship.
\end{acknowledgements}

\end{document}